\documentclass[12pt]{article}
\begin{document}
\begin{center}
{\bfseries
IS ISOSPIN ${\bf I=2}$ ASSIGNMENT PREFERABLE FOR ${\bf {d_1}^{*}(1956)}$ RESONANCE
CANDIDATE SEEN IN REACTION ${\bf PP\to 2\gamma X}$ BELOW PION THRESHOLD ?
}
\vskip 5mm
S.B.Gerasimov $^{1 }$ and M.Majewski$^{2}$
\vskip 5mm

{\small
(1) {\it Bogoliubov Laboratory of Theoretical Physics, JINR, Dubna} \\
(2) {\it University of Lodz, Department of Theoretical Physics, Lodz, Poland}\\
}
\end{center}
\vskip 5mm
\begin{center}

\begin{minipage}{150mm}
\centerline{\bf Abstract}
We compare the virtues of two options for the isospin ($I=1$ or $2$)
of the apparently exotic dibarion resonance $d_{1}^{*}(1956)$,
as it is noticed by the DIB2gamma Collaboration (JINR)
in the reaction $pp \to 2\gamma X$ below
the pion threshold, and using recent results
following from the RMC Collaboration (TRIUMF)
measurement of the relative probability of the double-photon
pion capture in the $\pi$-mesic deuterium atoms.
\vskip 5mm

\end{minipage}
\end{center}
\vskip 10mm

{\bf 1}.~~After obtaining the evidence of the excitation of the exotically
narrow dibaryon resonance with the mass around 1950 MeV,
 tentatively labelled as the $d_{1}^{\star}(1956)$,
in the reaction $pp \to 2\gamma X$~~below the pion threshold \cite{Dib01},
it is
imperative to question about its significance in different
reactions as a possible means of the confirmation and as a  source of the
necessary constraining information about the resonance quantum
numbers \cite{Ge98,Ge00,Ge01,Kh02,Kh03}.
In particular, attention was drawn \cite{Ge98,Ge00,Ge01} to
the radiative pion capture in pionic deuterium atoms, specifically
to the reaction channel $\pi^{-}d \to \gamma d_{1}^{\star} \to 2\gamma nn$,
where
the second stage of this reaction would not introduce additional
electromagnetic "smallness" to the total reaction matrix element due to
the presumed $BR(d_{1}^{\star}(1956) \to \gamma NN)\simeq 1$.
This would provide a very  spectacular enhancement of the doubly
radiative pion capture probability if the first stage
of the very resonance excitation really takes place and would not be
suppressed due to some reasons.

{\bf 2}.~~Recently, the RMC Collaboration at TRIUMF reported new data on the
branching ratio of the double-radiative capture of the $\pi^{-}$-meson
from atomic orbit(s) of the pionic deuterium atom \cite{Zol00}
and derived
the upper bound on the BR-value with the suggested narrow
dibaryon resonance in
the intermediate state of the capture reaction \cite{Tri03}.
Our aim is to demonstrate that a possible resonance effect in
the reaction considered depends strongly on a possible resonance isospin
and that one of the earlier suggested \cite{Ge00}  isospin values,
namely $I=2$, should result in a much
less important resonance reaction amplitude, seemingly not
in contradiction with the reported non-observation of the resonance effects
in Ref.\cite {Zol00,Tri03}.

We first recapitulate a few features of the $pp \to 2\gamma$ experiment
\cite{Dib01}. Two back-to-back photons produced in the proton-proton
interaction (with 216 MeV of the initial proton kinetic energy in
the lab. system) were registered in coincidence at 90$^{o}$
with respect to the initial proton momentum. The spectral distribution
of the registered photons exhibits the structure of two maxima,
the narrow one
was suggested to come from the narrow resonance production vertex while
the broad one results from the three-body decay of the same resonance.
The statistical significances for the narrow and broad peak are
5.3$\sigma$ and 3.5$\sigma$, respectively.
The apparent narrowness of the resonant structure lying below the pion
production threshold is a key feature
constraining the resonance quantum numbers to either $J^P$
($\it {i.e.}$, spin and parity) or isospin
$I$ such that the two-nucleon decay channel is forbidden either
strictly (by the Pauli principle) or approximately, but strongly enough,
due to the isospin selection rules ($\it {e.g.}$, if $I\geq 2$).
The electromagnetic mechanism of the production and decay of the
$d_1^{*}(\sim 1950)$-resonance implies the conservation of only
the third projection but not the total isospin value of
the initial state particles.
With the standard isospin properties of the electromagnetic current
operator, we have possible isospin values of the hadronic state
produced in collision
of two protons either $I=1$ or $I=2$. As a reasonable way to specify then
the relevant composite configuration structure of some general
($\it {i.e.}$ not explored yet) six-quark resonating state, we invoke the
decomposition of a six-quark state into the series over three-quark baryon
clusters within the $SU(6)\times O(3)$-symmetry basis states \cite{Har81}.
Confining ourselves by the lowest,in mass, configurations which are
decoupled from two-nucleon states,
one remains with the $\Delta N$-configuration as a leading one.
By $"\Delta"$ one can assume the three-quark cluster with
the quantum numbers of the real $\Delta(1232)$-resonance but with
a possible differently-distributed "effective" mass.
With the naturally chosen orbital moment $L=0$ for the ground state, we have
the set of possible quantum numbers $J^P=1^{+}$~and $2^{+}$ and $I=1$~and $2$.

{\bf 3.}~~Earlier \cite{Ge98}, this simple model was also employed for estimation of
the $\pi^{-}d \to \gamma d_{1}^{*}(M_{res}) \to 2\gamma nn$
relative probability under
the hypothesized $IJ^P=11^{+}$ set of quantum numbers.
Within the assumed model
and with $M_{res}=1956 \pm 6~MeV$\cite{Dib01} the resonance mechanism can be shown
to lead to $BR(\pi^{-}d \to \gamma d_{1}^{*}(M_{res}) \to 2\gamma nn)$
of the order of $10^{-3}$ which is about two orders of magnitude larger than
the preliminary value $1.6 \cdot 10^{-5}$~\cite{Zol00}.

At the same time, as a means of possible discrimination between two values
of the resonance isospin, $I=1$~or $I=2$,~~the important selectivity of the
measurement of the doubly radiative capture probability in the pionic
deuterium was suggested and emphasized in Refs.\cite{Ge00,Ge01}.

To discriminate between different values of the resonance isospin
options $(I=0$ $\it vs$ $I=2)$, one should study several
reactions differently sensitive to different isospin values.
To this end, it was proposed to make use of the double radiative
capture process  in the pionic deuterium in addition to the nucleon-nucleon
double bremsstrahlung reactions having in mind that the isoscalar deuteron
is the simplest weakly bound and most thoroughly investigated nucleus.
We need to know the matrix element of the electromagnetic transition
operator between the deuteron and the isovector two-neutron state.
In the threshold region of the radiative capture of
$\pi^{-}$ from one of the bound $nS$-atomic orbits
\cite {TRI81,TRI89}, the dominant contribution
to this matrix element is expected to coincide essentially with the
"seagull" Feynman graph leading to the Kroll-Ruderman(KR) low-energy theorem
for charged pion photoproduction on the isolated nucleon.
In this approximation,
the transition operator is easily seen to be transformed
as a component of the isovector under rotations
in the isospin space.  This means that, unlike the proton-proton double
bremsstrahlung reaction, the excitation of the narrow dibaryon
in the radiative pion capture on the isosinglet deuteron
should be suppressed strongly if the isospin of this resonance equals 2.
The explicit degree of this suppression depends, naturally, on
a specific features of the structure and dynamics of
the system and the reaction considered.
Keeping in mind our $N\Delta$ model, we mention a few possible
features of the reaction illustrating this expected suppression.
 In a deuteron,
 one of the bound nucleons should be transformed
into the (virtual) $\Delta$-resonance by either $\pi N\Delta$~-~or
$\gamma N\Delta$~-vertex.
In the first case, attaching the photon line
in all possible ways to the nucleon' or $\Delta$~-~lines we get a set
of Feynman diagrams giving the correction to the dominant KR
contribution in the cases where this dominant term is allowed, $\it {e.g.}$,
if $I(d_1^{*})$ would be equal $1$.
As is known \cite{Er66}, in the case of
the conventional $\pi^- + p\to n +\gamma$~-capture reaction,
the respective amplitude ratio, the
$["correction"/"KR-contribution"]$, is of the order of
$|\vec{q}|\omega/m_{\pi}^2$. In a deuteron,
$|\vec{q}| \simeq m_{\pi} \sqrt{\varepsilon /M_N}$, where $|\vec{q}|$ is
the bound pion momentum relative to the proton inside the deuteron,
$M_N$-the nucleon mass, $\varepsilon $~-~the deuteron binding energy.
Therefore, after squaring this small quantity, we would get
the correction of the order of $\sim 4\cdot 10^{-4}$
to the leading term, estimated in Ref.\cite{Ge98} under the
assumption of the isovector nature of the resonance $d_1^{\star}$.
In the case of the isotensor dibaryon resonance, the terms of the same order
would play the role of the main terms, as far as the KR term
will  be cancelled here, and we are left then with the resonance excitation
branching ratio of the order of $O(\sim 10^{-7}\div 10^{-6})$ which is markedly lower than
the total branching of the doubly radiative capture channel cited in
\cite{Zol00}.
There is, however, an additional decay channel if
one of the bound nucleons is transformed
into the (virtual) $\Delta$-resonance by the $\gamma N\Delta$~-vertex,
namely, the two-step transition with two intermediate neutrons on-mass-shell
\begin{equation}
 \pi^{-} + d \to 2n \to \gamma + d_1^{*},
\end{equation}
having no analog in the pion-nucleon radiative capture.

The nature of the suppression factor is different in this case.
The amplitude of the two-step reaction mechanism is represented by
a convolution of two amplitudes $\langle \pi^- d|2n \rangle$ and
$\langle 2n|\gamma d_1^{*}\rangle$. The first one is related to the measured
decay channel \\
 $\pi^{-} + d \to 2n$ while the second one is a partial
amplitude of the process $n+n \to \gamma + d_1^{*}$. The physical amplitude
of an analogous process $p+p \to \gamma + d_1^{*}$ has a different projection
of the isospin and includes all allowed partial waves $J^P$ in the initial
two-proton state while the intermediate two-neutron state has dominantly
$J^P=1^{-}$, as far as the capture of the stopped pions in the deuterium target
is known to proceed mainly from the $nS$~-~orbits, where $n=3,4$~
\cite{TRI89}.
So, our estimation of the suppression factor will be based on
our "effective" $N\Delta$~-~ model employed earlier for estimation
of the $d_1^{\star}$~-excitation in the $pp \to 2\gamma X$~-reaction
\cite{Ge98}.

The radiative $1^{-} \to \gamma + 1^{+}$~(or $2^{+}$)~transition with
the $\gamma N\Delta$~-~vertex refers to the spin-dependent electric-dipole
(E1) and magnetic-quadrupole (M2) (and/or electric-octupole, E3,
in case of $J^P=2^{+}$~-~option for the $d_1^{\star}$~-~resonance) amplitudes,
 and the expected suppression factor results from the appearance of
 a "retardation" factor $\sim(\vec{k}\cdot\vec{r})$ in the radial part
 of the total transition amplitude,; $\vec{r}$~being the distance between
 the initial neutrons having the kinetic energy $m_{\pi}/2=p^2/(2m_{N}$)
 each and described by the radial wave function $\sim j_1(pr)$ where
 the $j_1(x)$
 stands for spherical Bessel's function and $\vec{r}$ is also a relative distance
 vector between the nucleon and $\Delta$ composing the $d_{1}^{\star}$~
 resonance. Taking, for a semi-qualitative estimation of
 the "retardation" suppression factor K, the asymptotic form of
 the S-wave, bound $N\Delta$~ radial function,
 $R_b(r)\propto exp(-\alpha_{res} r)/r$, $\alpha_{res}\simeq
 \sqrt{2M_{red}\cdot \varepsilon_{res}}$, $M_{red}^{-1}=
 M_{\Delta}^{-1}+M_{N}^{-1}$, $\varepsilon_{res}\simeq M_n+M_{\Delta}-
 M_{res}$,
 the free-motion P-wave, radial nn-wave function,
 $R_{nn}(r)\propto j_{1}(pr)$,
 and averaging the result over the angles between the vectors $\vec k$ and
 $\vec p$ in the system where $\vec{p}/p=\vec{n}= \{0,0,1\}$, we get
\begin{equation}
K\simeq \frac{4\cdot \vert \int 3j_1(pr)(1/2)(\vec{k}\cdot \vec{r})
(\vec{n}\cdot \vec{r})R_b(r)d^{3}r\vert^2 }
{3\cdot \vert \int j_{0}(p^{\prime}r)R_b(r)d^{3}r \vert^2}
\approx 2.7\cdot \overline{cos^{2}\vartheta}\cdot10^{-3}
\sim {\cal O}(10^{-3})
\label{mod}
\end{equation}
where $p'=|\vec{p}-1/2 \vec{k}|\simeq p$ and we included in K the factors
of the averaging over possible initial spin values.
This very significant suppression factor brings earlier estimation
of the doubly radiative branching ratio due to the resonance excitation
down to the experimental uncertainty level of the RMC Collaboration
experiment.

{\bf 4.}~~Alternatively, one can proceed in a more qualitative but less model-dependent
way, invoking the DIB2$\gamma$  estimation of the experimental
$pp \to \gamma d_{1}^{\star}(1956)$~cross section and scaling it down to
the kinematics condition of the presumably resonant pionic capture
experiment.
We start with the inequality
\begin{equation}
|\langle \pi^{-}d|\gamma d_{1}^{*}\rangle| \leq (|\langle \pi^- d|2n \rangle|^2
|\langle 2n|\gamma d_1^{*}\rangle|^2)^{1/2}, \nonumber
\label{(1)}
\label{ineq}
\end{equation}
where summation and integration over quantum numbers and momenta of
the intermediate two-neutron states is understood.

Supplying (\ref{ineq}) with the kinematic factors needed to transform
$|\langle \pi^- d|2n \rangle|^2$~into \\
$\Gamma(\pi^- d \to 2n)$ and
$\langle 2n|\gamma d_1^{*}\rangle|^2$ into $\sigma_{tot}(n+n\vert_{J^P=1^-} \to
\gamma + d_1^{*}(1956))$, we get
\begin{eqnarray}
BR(\pi^- d \to \gamma d_{1}^{\star}(1956) \to 2\gamma 2n)\leq
BR(\pi^- d \to 2n)\cdot\frac{(W_{nn}^2-4m_n^2)}{16\pi}
\cdot\nonumber \\
(\frac{k(W_{nn})}{k(W_{pp})})^3\cdot
\sigma_{exp}(pp\to d_{1}^{\star}(1956))\times K_{max(mod)}
\simeq 9\cdot10^{-5}\times 1~(2\cdot10^{-3}).\label{bnd}
\end{eqnarray}
In (\ref{bnd}), the factors referring to the pion atomic wave functions
are mutually cancelled,
the $BR(\pi^- d \to 2n)\simeq .73$ is taken according to \cite{TRI81,TRI89},
$W_{nn}=m_d+m_{\pi}$, $W_{pp}=\sqrt{4m_p^2+2m_{p}T_p}$, with $T_p=216$~MeV,
and $\sigma_{exp}(pp\to \gamma d_{1}^{\star}(1956))\simeq (4\pi)^2\cdot 9~nb$
were taken from ref.\cite{Dib01},
the scaling factor $(k(W_{nn})/k(W_{pp}))^3$~~with
$k\equiv \omega=(W^2-m_{d^{\star}}^2/(2W)$~
is pertinent to the (minimal)
dipole multipolarity of the radiative $NN \to \gamma d_{1}^{\star}(1956)$
transition, the intermediate two-nucleon phase space and flux-factors
combine to produce other energy and mass dependent factor while
two versions of the residual scaling K-factor have the following meaning.
The $K_{max}\simeq 1$ would be valid in the idealized case
$\sigma (NN({J^P=1^{-}})\to \gamma d_{1}^{\star}(1956)) \approx
\sigma_{tot}(NN\to \gamma d_{1}^{\star}(1956))$ due to some
favourable spin-parity combination in the initial and final dibaryon
systems and appropriate dynamics of the radiative transition between them.
The model example, Eq.(\ref{mod}), is just the opposite case of
the unfavourable (or "hindered" transition) situation, where the
$NN(J^P=1^{-})$~-state is, presumably, transformed into the
$N\Delta(J^P=1^{+})$~-state via the dominantly magnetic-dipole
$\gamma N\Delta$~-coupling.

We note that if the scaling factor K is of the order of 1, then even in the case
of isotensor nature of the $d_1^{*}$~-resonance, the estimated value
$BR(\pi^{-}d \to 2\gamma 2n)\approx 10^{-4}$ turns out to be
$\sim 5$ times larger than
$BR_{exp}(\pi^{-}d \to 2\gamma 2n)\approx 1.6 \cdot 10^{-5}$~\cite{Zol00}.
Therefore, some kind of additional "hinderance",
{\it e.g.} of the type provided by our $N\Delta$ "toy"-model, is needed.
If the resonance quantum numbers are indeed $I=2, J^P=1^{+}$, one can foresee
one more experimentally interesting feature, namely, at the initial nucleon
kinetic energy $T_N\simeq 163$~MeV, the reaction $pn \to d_1^{*} \to pn$
can proceed via the really possible and measurable isospin-violating
mechanisms while the resonance formation in the $pp$- or $nn$~-
reactions will be strictly forbidden by the Pauli principle.

{\bf 5.}~~Returning to the very question of the existence and resonance
interpretation of the specific structure in the photon energy
distribution observed in the reaction $pp \to 2\gamma 2p$
below the pion threshold, we stress a real need to check, first of all,
the relation $\omega=(W^2-M_{d_1^{*}}^2)/(2W)$ between the photon energy and
mass of the resonance
at different values of W (or, equivalently, at different values of
the initial nucleon kinetic energy in lab.system), specifying
the excitation of the resonance with mass $M_{d_{1}^{\star}}$ at
the first stage of the reaction
$NN \to \gamma d_{1}^{\star} \to 2\gamma 2N$.
It was stated in
Ref.\cite{KVI99,KVI04} that during the course of investigation of the
reaction $pp \to e^{+}e^{-}2p$~ at the initial proton kinetic energy
$T_p=190$~Mev, a significant number of the two photon production events
was registered at the same energy of protons.
In the KVI-experiment, we would then expect the presence of
a narrower peak in the photon energy spectral distribution to be seen
at $\omega \simeq 13$~Mev while a broader
curve representing the energy distribution of the photon coming
from the 3-body resonance decay $d_{1}^{\star}(1956) \to \gamma 2p$,
would have a maximum around $\sim 50\div 60$~MeV,
as in the DIB2$\gamma$ experiment
carried out at $T_p\simeq 216$~MeV. Clearly, more detailed information
on other exclusive differential distributions in the final state of this
reaction would be of importance for the definition of the suggested
resonance quantum numbers.

{\bf 6.}~~To conclude, we have demonstrated that the reported
experimental results of
the search for the NN-decoupled (hence, very narrow) dibaryon resonance
in the $pp \to 2\gamma X$~ below the pion production threshold - with the
positive indication thereof \cite{Dib01} - and in the doubly radiative pion capture
process in deuterium mesoatoms - with no clear-cut resonance evidence
\cite{Zol00,Tri03} - can
be reconciled most easily if we assign the isospin $I=2$ to
the suggested $d_{1}^{\star}$~-resonance state with mass
$1956\pm 6$~MeV.
Furthermore, the spin-parity $J^P$ of this state should be such that
would provide further suppression of the partial cross section
$NN(J^P=1^{-}) \to \gamma d_{1}^{\star}$~through, {\it e.g.}, the operation
of a higher multipolarity mechanism in the considered radiative transition,
the magnetic quadrupole M2- or spin-dependent relativistic
correction to E1-~transition, to mention.

It seems relevant to notice, that if instead of the charged pion we would
take the neutral one participating in a given reaction, then
in the reactions including
the neutral pion, like $\gamma d \to \pi^{0} \gamma X \to 3\gamma X$,
we would expect more strongly enhanced exotic isotensor resonance excitation
and its subsequent radiative decay
$\gamma d \to \pi^{0} d_{1}^{*}(1956) \to \pi^{0} \gamma pn$
as compared to the nonresonance production of an "extra"-non-soft-photon
in the final state.
The detailed study ot the pion photoproduction reactions off the deuteron
as a meanse of looking for different kinds of the dibaryon exotics was carried
out in the recent work \cite{Fil04}.

Concerning the purposeful search for the $I=2$ dibaryon resonances in
pure hadronic reactions, the situation is controversial there as, in general,
in this type of the inquiry: there are both positive\cite{Tro00}
and negative\cite{Ta91} testimonies for that type of exotic hadron.
Especially interesting would be the states undergoing only weak decays.
The bounds on their existence, extracted from the experiments on
intermediate energy nuclear collisions ({\it e.g.}, Ref.\cite{Bo91}
and references therein) can depend rather critically on
their lifetime and, hence, on the resonance composition and internal
structure underlying the intensity of their interactions with the nuclear
environment and the resultant collisional broadening.

Finally, the chiral soliton model ($\chi$SM) predicting the exotically narrow
$\Theta^{+}$~-baryon \cite{Diak} shown up recently
with the mass value about 1540 MeV \cite{Nak03} would acquire also more
credibility in the two-baryon sector if (or, when) the reliably
reconfirmed $d_{1}^{\star}$ will display the characteristics
of long sought exotic dibaryons, theoretically suggested
or would-be explained within one or another approach based on $\chi$SM
( see, {\it e.g.},~\cite{Ko95} and references cited therein).

One of authors (S.G.) thanks Prof.J.Rembielinski
and the staff of the Department of Theoretical Physics
for warm hospitality and partial support during the visits to
University of Lodz.

Partial support of this work by the Bogoliubov-Infeld foundation is
gratefully acknowledged.

\end{document}